\documentclass[11pt,dvipsnames]{article}
\usepackage{times}
\usepackage{comment}
\usepackage{geometry}
\usepackage[utf8]{inputenc}
\usepackage{enumitem,amssymb}
\usepackage{multirow}
\usepackage{graphicx}
\usepackage{authblk}
\usepackage[backgroundcolor=white, bordercolor=white, textsize=tiny]{todonotes}
\usepackage{url}
\usepackage{xcolor}
\usepackage[
 colorlinks=true,    
 linkcolor=blue,     
 citecolor=blue,     
 filecolor=blue,  
 urlcolor=blue,      
 final=true
]{hyperref}
\usepackage{enumitem}
\setenumerate{itemsep=0mm}

\newcommand{\tbeta}{t_\beta}

\newcommand{\cba}{c_{\beta-\alpha}}

\newcommand{\mH}{m_H}
\newcommand{\mA}{m_A}
\newcommand{\mC}{m_{H^\pm}}

\newcommand{\bpa}{\textbf{BP-A}}
\newcommand{\bpb}{\textbf{BP-B}}

\newcommand{\be}{\begin{equation}\begin{aligned}}
\newcommand{\ee}{\end{aligned}\end{equation}}

\RequirePackage[normalem]{ulem}

\begin{document}

\title{
\vspace{-1cm}
\begin{flushright} \large DESY-22-090, KIAS--P22039 \end{flushright} 
\vspace{1cm}
\huge Snowmass 2021 \hfill \\[+1em]
\textit{Exotic Higgs Decays in the Type-II 2HDMs at Current and Future $pp$ Colliders} \hfill \\[+1em]
}
\date{} 

\author[a]{Felix Kling}
\author[b]{Honglei Li}
\author[c]{Shuailong Li}
\author[d]{Adarsh Pyarelal}
\author[e]{Huayang Song}
\author[c]{Shufang Su}
\author[f]{Wei Su}

\affil[a]{\small Deutsches Elektronen-Synchrotron DESY, Notkestr. 85, 22607 Hamburg, Germany}
\affil[b]{\small School of Physics and Technology, University of Jinan, Jinan, Shandong 250022, China }
\affil[c]{ \small Department of Physics, University of Arizona, Tucson, Arizona, USA 85721}
\affil[d]{\small School of Information, University of Arizona, Tucson, Arizona, USA 85721}
\affil[e]{\small \small CAS Key Laboratory of Theoretical Physics, Institute of Theoretical Physics, \\ Chinese Academy of Sciences, Beijing 100190, China}
\affil[f]{\small Korea Institute for Advanced Study, Seoul 02455, Korea}

\maketitle

\footnotetext[1]{\raggedright email addresses: \texttt{felix.kling@desy.de, shuailongli@email.arizona.edu, sps\_lihl@ujn.edu.cn, adarsh@arizona.edu, huayangs@itp.ac.cn, shufang@arizona.edu,weisu@kias.re.kr.}}
\noindent {\large \bf Thematic Areas:} \\
\noindent $\blacksquare$ (EF02) EW Physics: Higgs Boson as a portal to new physics \\ 
\noindent $\blacksquare$ (EF08) BSM: Model specific explorations \\
\noindent $\blacksquare$ (TF07) Collider Phenomenology \\

\noindent {\large \textbf{Abstract}: The exotic decay modes of non-Standard Model Higgses can serve as powerful search channels to explore the parameter space of extended Higgs sectors.  In this Snowmass contribution, we illustrate this using the Two-Higgs Doublet Model (2HDM) as an example. We first review the current experimental constraints on the parameter space of a Type-II 2HDM arising from existing searches for the exotic Higgs decay mode $A/H\rightarrow HZ/AZ$. We then present the sensitivity of future colliders to discover addition Higgs bosons using the exotic decay channels $A\rightarrow HZ$, $A\rightarrow H^\pm W^\mp$ and $H^\pm\rightarrow H W^\pm$. We find that a 100 TeV collider can probe  almost the entire region of the Type-II 2HDM parameter space that survives current theoretical and experimental constraints and would therefore be an ideal machine to search for heavier Higgses in hierarchical scalar sectors. 
}

\normalsize
\clearpage

\section{Introduction}
After the discovery of a light Standard Model (SM)-like Higgs boson at the LHC~\cite{Aad:2012tfa, Chatrchyan:2012xdj}, the search for Higgs `siblings' has become even more pressing. Two-Higgs-Doublet Models (2HDMs)~\cite{Branco:2011iw} comprise a well-motivated class of extensions to the SM Higgs sector. In addition to the SM-like Higgs, these models contain another CP-even Higgs $H$, one CP-odd Higgs $A$, and a pair of charged Higgses $H^\pm$.

Conventional searches for non-SM Higgses mainly focus on modes where they decay into pairs of SM particles. While these modes have been proven to be effective in the discovery of the SM-like Higgs, they suffer from the following limitations when it comes to searches for non-SM heavy Higgses:  (1) the couplings of $A$ and $H$ to the SM gauge bosons are suppressed since the current SM-like Higgs coupling measurements constrain the parameter region of 2HDMs to be close to the alignment limit, (2) the decay channels to pair of quarks have limited sensitivity due to large QCD backgrounds or non-trivial interference effects, and (3) the decay channels to pair of leptons are usually suppressed except in certain limited regions of parameter space.  

In the most general 2HDM, a \textit{hierarchical} spectrum -- that is, one whose states are sufficiently well-separated in mass, is still possible under both experimental and theoretical constraints~\cite{Kling:2018xud}.  Additional exotic decay channels with the decay of a heavy Higgs to a lighter Higgs and a SM gauge boson, or a heavy Higgs into two lighter Higgses become kinematically accessible.   Once these modes open up, they typically become dominant in large regions of parameter space, thus reducing the branching fractions of the conventional decay modes and relaxing the experimental limits based on them.  More importantly, these exotic decay channels can serve as new discovery modes for the non-SM heavy Higgses.  Detailed collider studies of various exotic decay modes and their reach at current and future $pp$ colldiers can be found in Refs.~\cite{Kling:2018xud, Li:2020hao, Kling:2016opi,Kling:2020hmi, Kling:2016opi,Kling:2015uba,Li:2015lra,Coleppa:2014cca,Coleppa:2014hxa,Coleppa:2013xfa,Hajer:2015eoa}. In this Snowmass whitepaper, we summarize the main results of those previous studies.

\section{Hierarchical Two Higgs Doublet Models and Exotic Higgs Decays}

The Higgs sector of the 2HDM consists of two SU(2)$_L$ scalar doublets $\Phi_i\,(i=1,2)$ with hyper-charge $Y=1/2$. After electroweak symmetry breaking, the neutral components $\Phi_{1,2}$ acquire vacuum expectation values of $v_{1,2}$. For convenience, we parameterize the 2HDM by the physical Higgs masses ($m_h, m_H, m_A$ and $m_{H^\pm}$), the mixing angle between the two CP-even Higgses ($\alpha$), the ratio of the two vacuum expectation values ($\tbeta=v_2/v_1$)  and the $\mathcal{Z}_2$ symmetry breaking parameter $m_{12}^2$. In the following, we will identify $h$ as the SM-like $125$ GeV Higgs, and the \textit{alignment limit} is therefore $\cba=0$.  In our study, we focused on the Type-II 2HDM, in which $\Phi_1$ couples to the charged leptons and down-type quarks and $\Phi_2$ couples to the up-type quarks.

Given the theoretical and current experimental constraints on the parameter space of the Type-II 2HDM, we follow Ref.~\cite{Kling:2016opi} and adopt two benchmark planes for collider studies on these exotic decay channels: 
\begin{itemize}
    \item{\bpa}  corresponding to the mass ordering $m_{A}>m_{H}=m_{H^\pm}$, permitting the exotic Higgs decay channels $A\rightarrow HZ\rightarrow (bb/\tau\tau/t t)\ell\ell$ and $A\rightarrow H^\pm W^\mp\rightarrow t b\ell\nu$.
    \item{\bpb}  corresponding to the mass ordering $m_{A}=m_{H^\pm}>m_{H}$, permitting the exotic Higgs decay channels $A\rightarrow HZ\rightarrow (bb/\tau\tau/t t)\ell\ell$ and $H^\pm\rightarrow H W^\pm\rightarrow (\tau\tau/tt) W^{\pm}$. 
\end{itemize}


\section{Current Constraints from the LHC}
\begin{figure}[htb]
\centering
\includegraphics[width=0.48\textwidth]{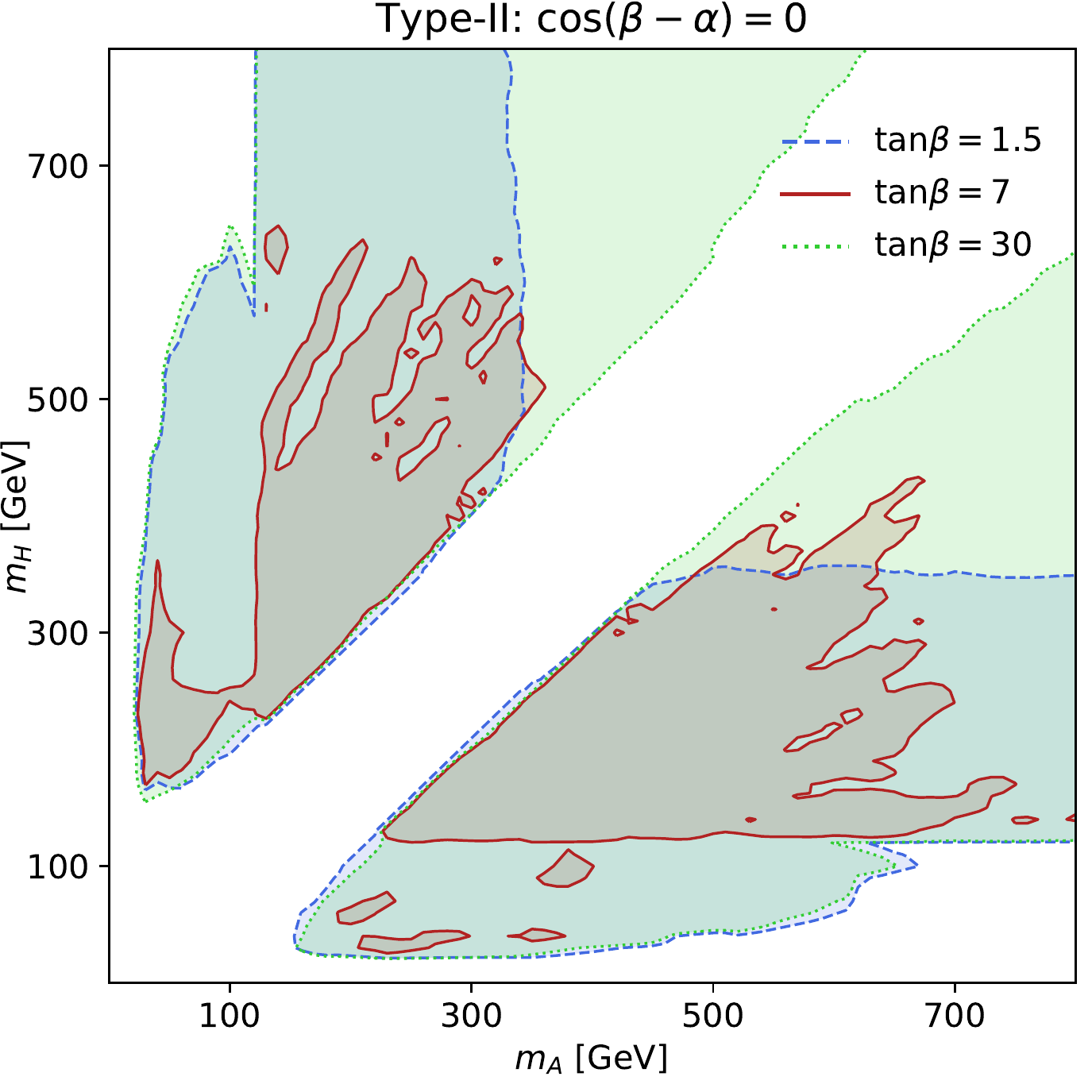}
\includegraphics[width=0.48\textwidth]{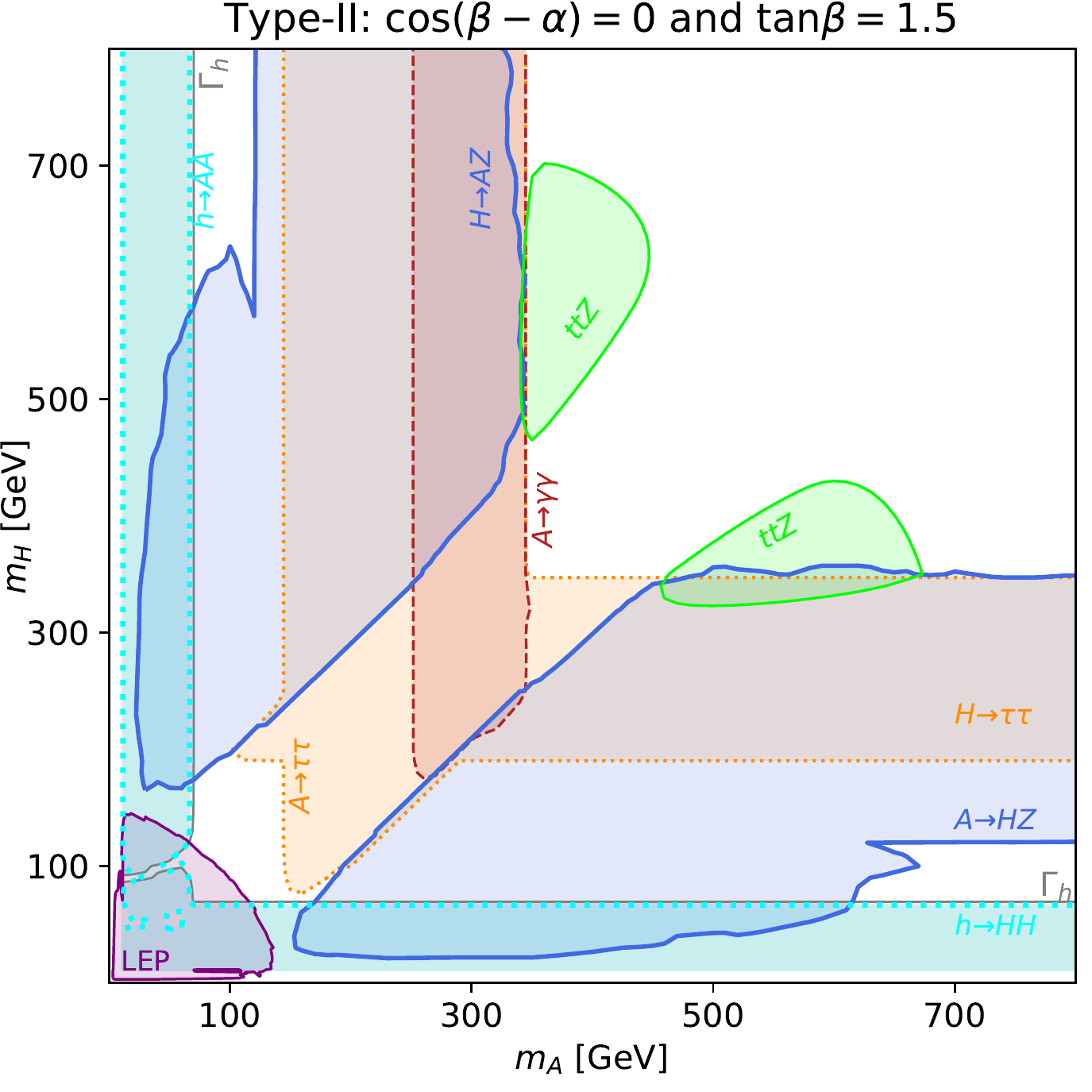}
\caption{Constraints excluded at $95\%$ C.L. by the $A/H\rightarrow HZ/AZ$ search in the alignment limit, $\cba=0$, on the Type-II 2HDM in $m_A$ vs. $m_H$ plane for different values of $\tbeta$ (left panel), and global constraints at $95\%$ C.L. for $\cba=0$ and $\tbeta=1.5$ from LHC searches for different experimental search channels (right panel)~\cite{Kling:2020hmi}.}
\label{fig:LHC_limit}
\end{figure}

Both ATLAS and CMS have already performed analyses looking for the most promising final states of the exotic decay channels of $A/H\rightarrow HZ/AZ\rightarrow (bb/\tau\tau)\ell\ell$~\cite{ATLAS:2018oht, CMS:2019ogx, CMS:2016xnc}. We re-interpret the experimental results and show the  limits on the 2HDM parameter space~\cite{Kling:2020hmi}. In the left panel of Fig.~\ref{fig:LHC_limit}, we show the constraint from the $A/H\rightarrow HZ/AZ$ channel for the Type-II 2HDM in the alignment limit $(\cba=0)$ for various $t_\beta$ value.   While at low $\tbeta=1.5$, the $13$ TeV searches exclude parent particle masses up to $800$ GeV for a daughter particle mass between $80$ and $350$ GeV.  At large $\tbeta{}$, the Type-II 2HDM has an enhanced reach due to the $\tbeta$ enhancement of bottom (and $\tau$) Yukawa couplings.

In the right panel, we combine the search limits from  $A/H\rightarrow HZ/AZ$ for $t_\beta=1.5$ with  constraints obtained from other search channels under the alignment limit.  The combination of all channels covers the majority of the region where one of the Higgs masses is below the di-top threshold, $m_{A, H}<2m_t$. The gap region between the measurements of the Higgs width, $\Gamma_h$, and conventional final state searches, $A/H\rightarrow \gamma\gamma/\tau\tau$ is mostly covered by the exotic Higgs decay modes $A/H\rightarrow HZ/AZ$, which have unique sensitivity to the alignment limit region.

\section{Exotic Higgs Decays at Future Colliders}
A future $pp$ collider, like the Future Circular Collider (FCC) at CERN~\cite{Abada:2019lih, Benedikt:2018csr} or the Super proton-proton Collider (SppC) in China~\cite{CEPC-SPPCStudyGroup:2015csa},  with a center of mass energy around 100 TeV, would be an ideal machine to study heavy non-SM Higgses.  At such a machine, top quarks produced in heavy particle decays will be highly boosted, resulting in fat jets that can be effectively identified using top-tagging techniques and distinguished them from the large SM backgrounds that typically pose a challenge at the LHC.

\begin{figure}[htb]
\centering
\includegraphics[width=0.48\textwidth]{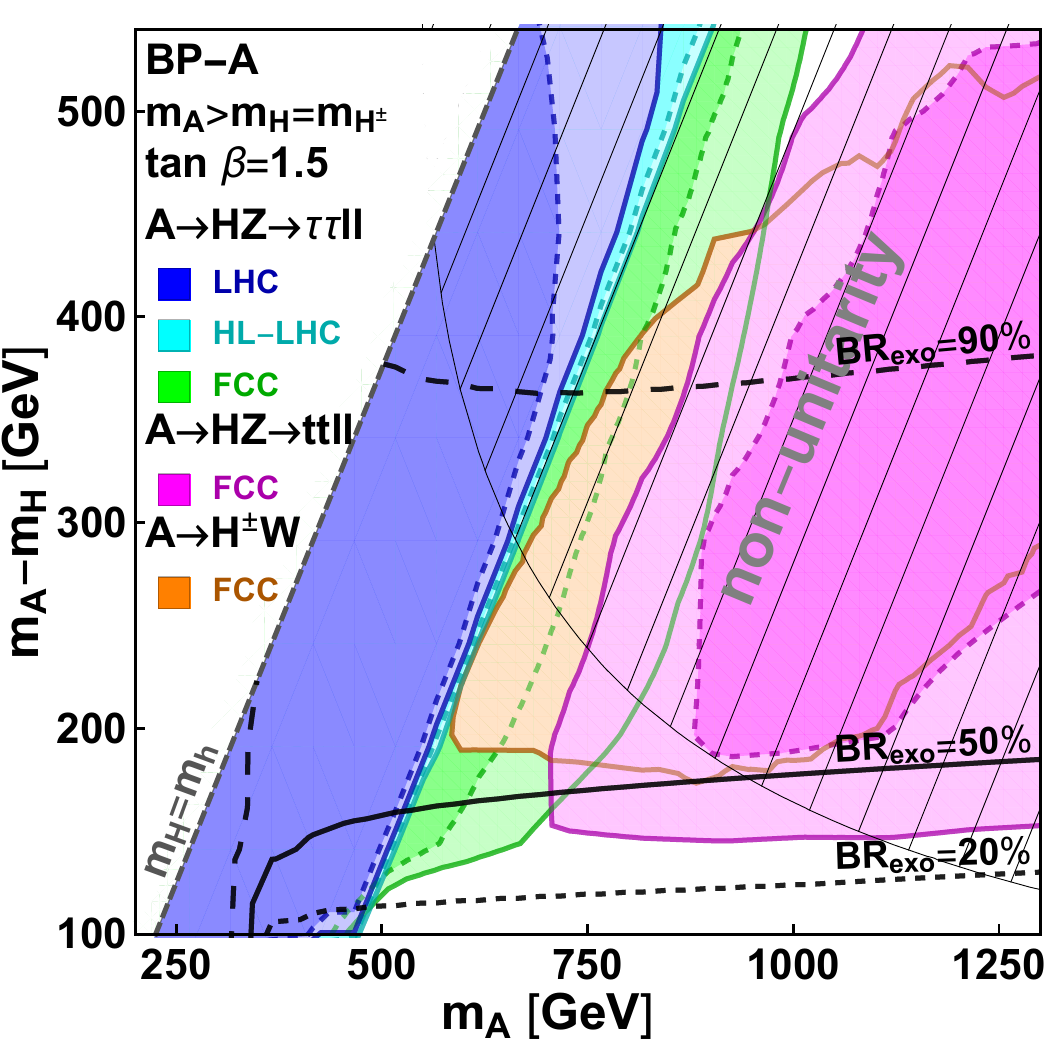}
\includegraphics[width=0.48\textwidth]{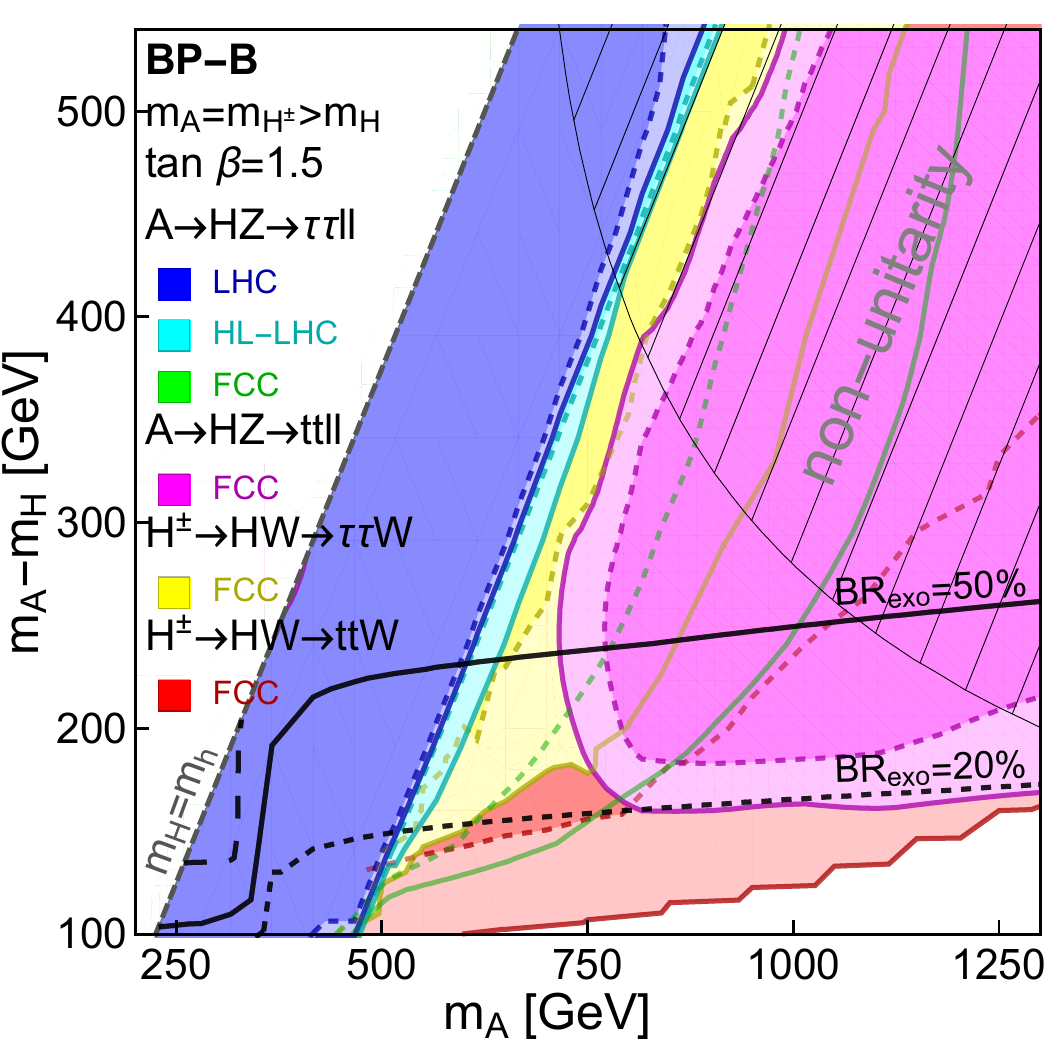}
\caption{Reach for the exotic Higgs decay channels at the LHC, HL-LHC and 100 TeV $pp$ collider for \bpa{} (left) with the mass hierarchy $\mH=\mC<\mA$ and \bpb{} (right) with the mass hierarchy $\mH<\mC=\mA$~\cite{Kling:2018xud, Li:2020hao}.  The dashed and solid curves indicate the 5$\sigma$ discovery and 95\% C.L. exclusion region, respectively.
} 
\label{fig:BP-A-B-tb1.5}
\end{figure}

We perform a detailed collider analysis to
obtain the $95\%$ C.L. exclusion limits as well as $5\sigma$ discovery reach at a future 100 TeV hadron collider for benchmark planes \bpa{} and \bpb{} by using state-of-the-art machine learning and top-tagging techniques~\cite{Kling:2018xud, Li:2020hao}. In Fig.~\ref{fig:BP-A-B-tb1.5}, we present the exclusion and discovery reaches in  the $\Delta m=m_A-m_H$ versus $m_A$ plane for \bpa~(left panel) and \bpb~(right panel) with $\tbeta=1.5$. The choice of the value of $\tbeta$ is representative of the interesting low $\tan\beta$ region, which is particularly hard to constrain using the conventional 
search channels such as $A/H \to \tau\tau$ and $H^\pm \to\tau \nu$ which are expected to provide the best sensitivity at higher values of $\tbeta$. 

We find that the highest sensitivity at low values of $m_A$ is provided by the $A \to HZ \to \tau\tau\ell\ell$ channel ({\color{blue} blue} for the LHC, {\color{cyan} cyan} for the HL-LHC, and {\color{green}green} for the FCC).
Once the $H\to tt$ channel is kinematically accessible, the decay channel $A \to HZ \to tt\ell\ell$ ({\color{Magenta} magenta}) becomes powerful to search for larger Higgs masses $m_A$ at a 100 TeV hadron collider. \bpa~also permits the additional exotic Higgs decay channel $A \to H^\pm W$ ({\color{Orange} orange}), and \bpb~provides the exotic decays of a charged Higgs $H^{\pm}\rightarrow HW^{\pm}\rightarrow (\tau\tau/tt) W^{\pm}$ ({\color{Yellow} yellow} and {\color{Red} red}). All these channels are reachable at a 100 TeV collider and complement each other nicely. Combining the aforementioned exotic decay channels, we find that almost the entire parameter space in which the exotic decay branching fraction is more than $\sim20\%$ can be probed.   We also note that the decay channels to $t\bar{t}$ are quite effective for heavy scalars in the small $\tan\beta$ region and complementary to the traditional $bb/\tau\tau$ search states.

\section{Summary}
While most searches for additional Higgs bosons have focused on conventional decay channels, we showed that, in general 2HDMs with a hierarchical mass spectrum, exotic Higgs decay channels are possible and can significantly reduce the sensitivity of the conventional decay searches once kinematically accessible. Furthermore, they serve an alternative discovery channels for the non-SM heavy Higgses.  We summarized the current experimental direct search limits on the parameter space of the Type-II 2HDM, in particular, including the exotic Higgs decay mode of $A/H\rightarrow HZ/AZ\rightarrow (bb/\tau\tau)\ell\ell$.   We also performed a multivariate analysis for the collider reach of various exotic Higgs decay channels at a 100 TeV $pp$ collider. These channels show significant sensitivities at future colliders. They are not only complementary to the conventional searches, but also complementary to each other. Combining all these search channels, almost the entire parameter space in hierarchical Type-II 2HDMs (i.e, with mass splittings $m_A-m_H\gtrsim 150$ GeV) can be explored at a future 100 TeV hadron collider.

\bibliography{references}

\end{document}